# The p-filter: multi-layer FDR control for grouped hypotheses

Rina Foygel Barber and Aaditya Ramdas

October 28, 2016

**Abstract**

In many practical applications of multiple testing, there are natural ways to partition the hypotheses into groups using the structural, spatial or temporal relatedness of the hypotheses, and this prior knowledge is not used in the classical Benjamini-Hochberg (BH) procedure for controlling the false discovery rate (FDR). When one can define (possibly several) such partitions, it may be desirable to control the *group-FDR* simultaneously for all partitions (as special cases, the "finest" partition divides the $n$ hypotheses into $n$ groups of one hypothesis each, and this corresponds to controlling the usual notion of FDR, while the "coarsest" partition puts all $n$ hypotheses into a single group, and this corresponds to testing the global null hypothesis).

In this paper, we introduce the *p-filter*, which takes as input a list of $n$ p-values and $M \geq 1$ partitions of hypotheses, and produces as output a list of $\leq n$ discoveries such that group-FDR is provably *simultaneously* controlled for all partitions. Importantly, since the partitions are arbitrary, our procedure can also handle multiple partitions which are nonhierarchical. The p-filter generalizes two classical procedures—when $M = 1$, choosing the finest partition into $n$ singletons, we exactly recover the BH procedure, while choosing instead the coarsest partition with a single group of size $n$, we exactly recover the Simes test for the global null. We verify our findings with simulations that show how this technique can not only lead to the aforementioned multi-layer FDR control, but also lead to improved *precision* of rejected hypotheses. We present some illustrative results from an application to a neuroscience problem with fMRI data, where hypotheses are explicitly grouped together according to predefined regions of interest (ROIs) in the brain, thus allowing the scientist to explicitly and flexibly employ field-specific prior knowledge.

## 1 Introduction

One of the biggest concerns in the reproducibility crisis faced by modern data analysis is the practice of testing hundreds or thousands of hypotheses often arising from a single experiment. One of the earliest methods to gain some control on the number of false discoveries (null hypotheses that were incorrectly rejected by the scientist) is the Bonferroni correction, which controls the *family-wise error rate* (FWER), which requires that the probability of making any false discoveries must be bounded by $\alpha$. This procedure, which compares each p-value against the "corrected" threshold $\frac{\alpha}{n}$ (where $n$ is the number of hypotheses), is known to lead to extremely low power. Since then, a wide range of methods have been proposed as alternatives, such as the test by [12] for the "global null" (testing whether all hypotheses are null). Closely related to Simes' test, the most practically popular method is the procedure by [2] (BH) for controlling the *False Discovery Rate* (FDR).

We refer to "true signals" to mean those tests for which the null hypothesis is actually false (and should be rejected), and "nulls" or "true nulls" to mean those tests with no real signal, where the null hypothesis is true (and should not be rejected). Our "discoveries" are those tests which our method identifies as likely true signals (i.e., our algorithm's rejected null hypotheses). A false discovery is, of course, a false rejection: a null hypothesis that was rejected by our algorithm (proclaimed as a discovery) but is in fact a true null hypothesis.

We propose an algorithm called the p-filter, which is an elegant conceptual unification and generalization of the BH procedure and Simes' test for the global null, which is useful in practical scenarios when the scientist can naturally partition the hypotheses being tested into groups, and desires to control both the *overall FDR* (controlling the number of falsely discovered hypotheses) and the *group FDR* (controlling the number of falsely discovered groups). We say that a group is said to be falsely discovered if there is at least one hypothesis rejected within that group, but in reality the group consists entirely of nulls. Our procedure can also handle multiple partitions, referred to as "layers", which *are not necessarily required to be hierarchical*; the p-filter provides FDR control simultaneously at the level of each specified "layer". Practitioners may use prior knowledge to group together hypotheses that they expect to be either simultaneously false or simultaneously true, or organize the hypotheses according to some discipline-specific natural partitioning. At a high level, the p-filter works by filtering the groups in each partition, or "layer", searching for groups that pass some threshold of evidence for a true signal; in the end, a hypothesis is rejected if and only if it passes through every layer of the filter.

Consider an example from neuroscience where controlling FDR is both crucial and already popular since the early adoption popularized by [6]. Consider showing a patient some stimulus and recording some physiological correlate of her brain activity (using, say, fMRI). Suppose we consider brain locations (voxels) $z_1, ..., z_V$, at times $t_1, ..., t_S$ after presentation of the stimulus, and formulate the following $V \cdot S$ many null hypotheses:

$$\mathrm{H}^0_{(v,s)} : \text{The stimulus is independent of activity at } v\text{, at delay } s \text{ after presentation.}$$

In addition to controlling the usual FDR using the trivial partition (treating each $(v, s)$ as its own group), we may want to ensure that the group-FDR is *also* simultaneously small, where one may partition the hypotheses into voxels (grouping $(v, s)$ for fixed $v$, across all delays $s$) and/or into timepoints (grouping $(v, s)$ for fixed $s$, across all voxels $v$ in some functional region). Note that in this example, the three layers are not hierarchical—when we partition by space and by time, neither partition can be nested inside the other.

Another area where such groupings may be natural is bioinformatics or statistical genetics – when looking for associations between genes and proteins, it may make sense to group together proteins with similar amino-acid structure, and/or group together genes with similar nucleic acid sequences, perhaps employing prior knowledge from existing gene ontologies. We also expect our work to find favor in other spatio-temporal applications of FDR, whenever rejected hypotheses are expected to be contiguous in space and/or time.

**Related work** The nearest comparison to our method in the literature is the work of [1] who proposed a hierarchical FDR control procedure, developed further by [10]. We return to this work in Section 3, where we discuss group-wise FDR control, and again in Section 6, where we compare our method to theirs conceptually and empirically. [16] also considers the problem of testing hypotheses which are arranged in a hierarchy; this is of course related to simultaneously controlling group-wise and overall FDR. Our procedure is more general than both of these methods since, for the p-filter, the various layers/partitions are not required to form a hierarchy.

Other recent papers have examined related questions. Here we briefly describe several such works, but many more exist in the literature. In the variable selection problem for a regression framework, [9] considers hierarchical tests for handling clusters of highly correlated variables. A different setting also involving grouped hypotheses arises in [7], where the goal is to control the overall FDR only, but different groups of hypotheses have different proportions of true signals vs. nulls; by estimating these proportions for each group separately, their method increases power to detect true signals in the high-signal groups. Hypotheses may be grouped in a data-dependent or adaptive way in some applications, for example in spatial data where locally contiguous regions can form a "cluster" of discoveries; the problem of controlling false discoveries at the cluster level is studied by [4] and [13].

**Outline** In Section 2, we recall various standard definitions and the standard FDR procedure of [2]. Next, we present our method; for the purposes of clarity, we split our exposition into two parts. In Section 3, we show how to control FDR simultaneously for individual hypotheses and at the group level, if our set of hypotheses is partitioned into groups. This leads into the more general setting of Section 4, where we develop the p-filter for controlling FDR across an arbitrary number of (possibly non-nested) partitions, or "layers". An algorithm for running the p-filter efficiently is given in Section 5. We then examine the empirical performance of our method on simulated data in Section 6 and on fMRI data in Section 7. We give some concluding remarks in Section 8. Proofs for our theoretical results are deferred to Appendix A.

# 2 Background

We assume the reader is familiar with the classical setup of frequentist hypothesis testing. In this paper, we assume that we are given a set of p-values, denoted by the vector $P \in [0,1]^n$, each corresponding to a different question (a different null hypothesis), and we wish to select some subset of these tests as our "discoveries" (i.e., to reject some subset of the corresponding null hypotheses) while retaining some form of control over the number of false discoveries. For the remainder of the paper, let $\mathcal{H}^0 \subseteq [n]$ be the set of tests (hypotheses) designated as "true nulls", and let $\widehat{S}$ be the set selected as our discoveries based on the observed p-values.

## 2.1 Benjamini-Hochberg procedure for FDR control

For an algorithm that chooses a set of hypotheses to reject (denoted here by $\widehat{S}$), the seminal paper by [2] proposed to measure its performance via the *False Discovery Rate (FDR)*, defined as

$$\text{FDR} = \mathbb{E}\left[\frac{|\mathcal{H}^0 \cap \widehat{S}|}{1 \vee |\widehat{S}|}\right]$$

where $|\mathcal{H}^0 \cap \widehat{S}|$ is number of false discoveries (null hypotheses that are true, and are incorrectly rejected) and $|\widehat{S}|$ is the total number of discoveries (all hypotheses that are rejected). The notation $1 \vee |\widehat{S}|$ in the denominator is defined as $\max\{1, |\widehat{S}|\}$ and ensures that, if no rejections are made, then the false discovery proportion is defined as zero.

Given the vector of p-values $P = (P_1, \ldots, P_n)$, the Benjamini-Hochberg (BH) procedure with target FDR level $\alpha$ is defined by calculating

$$\widehat{k}_\alpha(P) = \max\left\{k \in \{1, \ldots, n\} : \left|\left\{i : P_i \le \frac{\alpha \cdot k}{n}\right\}\right| \ge k\right\} \ ,$$

with the convention that we set $\widehat{k}_\alpha(P) = 0$ if this set is empty. Equivalently, if $P_{(i)}$ is the $i$th smallest p-value, then

$$\widehat{k}_\alpha(P) = \max\left\{k \in \{1, \ldots, n\} : P_{(k)} \le \frac{\alpha \cdot k}{n}\right\} \ ,$$

The method then rejects the $\widehat{k}_\alpha(P)$ smallest p-values, or equivalently, rejects all p-values that are $\le \frac{\alpha \cdot \widehat{k}_\alpha(P)}{n}$. The authors then showed that this procedure provably controls the FDR at level $\alpha$ if the p-values are independent. Subsequent work by [3] proved that this result holds under a relaxed condition, the PRDS assumption (Positive Regression Dependence on a Subset), where the p-values are allowed to have positive dependence (see Eq. (5) below for details).

## 2.2 Simes test for the global null

The BH procedure is closely related to earlier work by [12], which for a vector of p-values $P = (P_1, \ldots, P_n)$, tests the *global null* hypothesis (also called the intersection hypothesis), that is, tests whether *all* of these $n$ p-values are null (there are no true signals). To perform this test, first calculate the Simes p-value

$$\text{Simes}(P) = \min_{1 \le k \le n} \frac{P_{(k)} \cdot n}{k},$$

where as before, $P_{(k)}$ is the $k$th smallest p-value in the list $P_1, \ldots, P_n$. The global null hypothesis is then rejected if $\text{Simes}(P) \le \alpha$, where $\alpha$ is the prespecified level of the test (the desired Type I error rate).

To see the connection to the BH procedure, for any $n \ge 1$ and $\alpha \in [0, 1]$, write

$$P \in \text{BH}(\alpha)$$

whenever $\widehat{k}_\alpha(P) \ge 1$, that is, this is equivalent to the statement that the set of p-values $P$ leads to at least one rejection, when applying the BH procedure with target FDR level $\alpha$. We then say $P$ *passes* BH at level $\alpha$. Examining the definition of the BH procedure, we see that

$$\text{Simes}(P) = \min\left\{\alpha \in [0, 1] : P \in \text{BH}(\alpha)\right\} \ ,$$

that is, the Simes p-value is the minimum threshold $\alpha$ for which $P$ passes the BH procedure. In other words,

$$\text{Simes}(P) \le t \ \Leftrightarrow \ P \in \text{BH}(t) \ \Leftrightarrow \ \widehat{k}_t(P) \ge 1 \tag{1}$$

for any $t \in [0, 1]$. We should note that the Simes p-value really is a p-value in the true sense of the word—if the p-values are independent and uniform, then the Simes p-value is uniformly distributed under the global null (i.e., if $P_1, ..., P_n$ are independent and uniformly distributed). This is because

$$\mathbb{P}\left\{\text{Simes}(P) \le t\right\} = \mathbb{P}\left\{P \in \text{BH}(t)\right\} = t$$

where the latter equality is a property of BH under the global null [2]. Under positive dependence (i.e., PRDS), the Simes p-value becomes conservative, with $\mathbb{P}\left\{\text{Simes}(P) \le t\right\} \le t$ by properties of BH under positive dependence [3].

## 2.3 FDR control *only* at the group level: interpolating between Simes & BH

Suppose for a moment that all the p-values are independent and uniformly distributed, and that we have partitioned our hypotheses into $G$ groups of size $n_1, n_2, \ldots, n_G$, with $n = n_1 + \cdots + n_G$:

$$\underbrace{P_1, \ldots, P_{n_1}}_{\text{Group 1}}, \underbrace{P_{n_1+1}, \ldots, P_{n_1+n_2}}_{\text{Group 2}}, \ldots, \underbrace{P_{n_1+\cdots+n_{G-1}+1}, \ldots, P_n}_{\text{Group } G},$$

and we wish to select a subset of these groups, $\widehat{S}_{\text{grp}} \subseteq [G]$, so that the proportion of null groups is not too high. (In this setting, a "null group" is a group consisting entirely of null hypotheses.)

We can consider the following simple procedure for this problem. Using the Simes p-value, we could reduce this to a standard multiple testing problem: specifically, we compute the Simes p-values for each of the $G$ groups,

$$\text{Simes}(P_{A_1}), \ldots, \text{Simes}(P_{A_G}),$$

where $A_g = \{n_1 + \cdots + n_{g-1} + 1, \ldots, n_1 + \cdots + n_g\}$ is the set of indices belonging to group $g$, and $P_{A_g}$ is the vector of p-values belonging to this group. Then, apply the BH procedure with threshold $\alpha$ to this new list of p-values to produce a set $\widehat{S}_{\text{grp}}$ of (group) discoveries. If the $n$ p-values are independent, then since we are simply applying BH to a set of p-values (which are independent and, for each null group, are uniformly distributed), we can then expect this procedure to control group-level FDR, and indeed it immediately follows that

$$\mathbb{E}\left[\frac{|\mathcal{H}^0_{\text{grp}} \cap \widehat{S}_{\text{grp}}|}{1 \vee |\widehat{S}_{\text{grp}}|}\right] \leq \alpha.$$

(Of course, we would have no corresponding guarantee for the overall FDR when the hypotheses are considered individually rather than in groups; our multilayer method, introduced shortly, gives this type of simultaneous guarantee.)

In fact, we can view this type of group FDR procedure as an interpolation between the Simes test of the global null, and the Benjamini-Hochberg procedure. That is, both the Simes test and the Benjamini-Hochberg procedure are actually special cases of the group-FDR-control method described in this section, obtained by considering two extremes: one group of size $n$ (corresponding to the Simes test of the global null), or $n$ groups of size one (corresponding to the BH procedure). It is intuitively pleasing that our multilayer method, to be introduced later, also specializes to the Simes test and the BH procedure in the case of only one layer, exactly in the fashion mentioned above.

## 2.4 Independent group and individual level discoveries may conflict

Unfortunately, for two (or more) layers, controlling FDR both at the group level (using the Simes+BH procedure in the previous subsection) and independently at the individual level (using the BH procedure) may cause conflicts in rejected groups and individual hypotheses. The example in Figure 1 is meant to demonstrate exactly this issue, highlighting the complications that may arise in the multilayer setting, even for just two layers. Here, we divide 20 p-values into 4 groups of 5 p-values each, and choose to control the FDR at the individual and group levels, both at $\alpha = 0.2$.

The row and individual level rejections in Figure 1 are in conflict, because the third row is discovered at the group level but does not contain any hypotheses discovered at the individual level; conversely, the fourth row was not discovered at the group level, but has a p-value discovered at the individual level. Thus while these outcomes guarantee FDR control at the group and individual levels, the output of this procedure is not *internally consistent*. If we throw away all rejections that are in conflict, by rejecting the individual hypotheses only from the first two rows (i.e. taking the intersection of rejections at different layers) and discarding the group rejection of the third row and the individual rejection in the fourth row, we now have a result that is internally consistent, but unfortunately we have lost the guarantees of FDR control—indeed, it is a well known property of BH that rejecting fewer hypotheses than recommended by the BH procedure may sometimes increase the FDR.

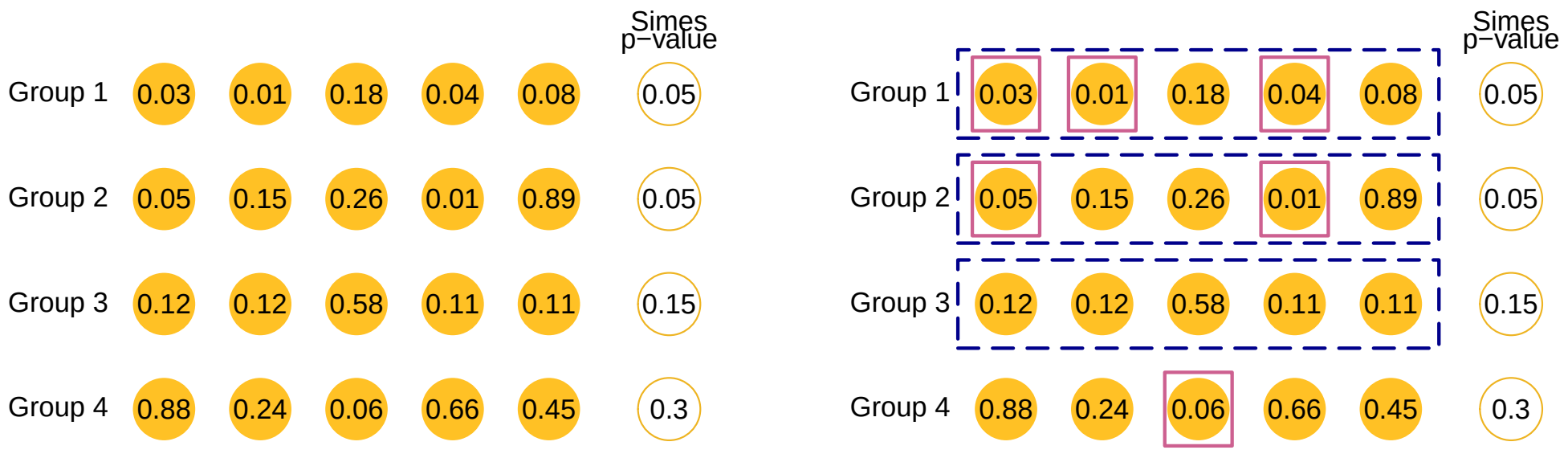

Figure 1: On the left, 20 p-values and their groupings (into rows) are displayed in the shaded circles, and the Simes p-values for the groups are displayed in the unshaded circles. On the right, the discoveries made by running the BH procedure on the 20 p-values, with $\alpha = 0.2$, are portrayed by solid-line squares, and the discoveries made by running the group-FDR controlling procedure independently (BH applied to the Simes p-values for each group), with $\alpha = 0.2$, are portrayed by the dashed-line rectangles.

There are special cases where the group and individual level rejections may not be in conflict, as discussed in [10]. However, this certainly does not generalize to arbitrarily many layers of arbitrary groups being tested at arbitrary levels. This motivates the further study of procedures that can provide *simultaneous* FDR guarantees for multiple possibly non-hierarchical layers. Indeed, our general and efficient p-filter algorithm provably gets around the obstacles mentioned above in quite some generality.

## 3 Controlling FDR for individual hypotheses and for groups

Assume again that we partition our set of $n$ hypotheses (and their corresponding p-values) into $G$ groups of size $n_1, n_2, \dots, n_G$:

$$\underbrace{P_1, \dots, P_{n_1}}_{\text{Group 1}}, \underbrace{P_{n_1+1}, \dots, P_{n_1+n_2}}_{\text{Group 2}}, \dots, \underbrace{P_{n_1+\dots+n_{G-1}+1}, \dots, P_n}_{\text{Group } G},$$

with $n = n_1 + \dots + n_G$. Let $\mathcal{H}^0 \subseteq [n]$ index the unknown set of null hypotheses, and define the set of null groups (groups that contain only null hypotheses) as

$$\mathcal{H}^0_{\text{grp}} = \{g : A_g \subseteq \mathcal{H}^0\}$$

where $A_g$ is the set of indices belonging to group $g$ as before. We now consider the problem of controlling FDR at the individual and the group level simultaneously, possibly for different target FDR levels $\alpha_{\text{ov}}, \alpha_{\text{grp}}$.

### 3.1 Overall FDR and group-level FDR

We now present our proposed method, the p-filter, for controlling FDR at both granularities, i.e., the standard overall FDR and the group level FDR.

First, consider a pair of thresholds $(t_{\text{ov}}, t_{\text{grp}}) \in [0,1] \times [0,1]$ (we show below how p-filter chooses these thresholds adaptively; for the purposes of definitions assume they are given). At this pair of thresholds, we define the set of all "discoveries" (rejections) made by the algorithm as

$$\widehat{S} = \widehat{S}(t_{\text{ov}}, t_{\text{grp}}) = \left\{ i : P_i \leq t_{\text{ov}} \text{ and } \text{Simes}(P_{A_{g(i)}}) \leq t_{\text{grp}} \right\} , \tag{2}$$

where $g(i)$ is the group to which $P_i$ belongs. In other words, a hypothesis is rejected if and only if its p-value $P_i$ is below the overall threshold $t_{\text{ov}}$ *and* the Simes p-value for its group, $P_{A_{g(i)}}$, is below the group threshold $t_{\text{grp}}$. Next, define the set of group discoveries as

$$\widehat{S}_{\text{grp}} = \widehat{S}_{\text{grp}}(t_{\text{ov}}, t_{\text{grp}}) = \{ g : \widehat{S}(t_{\text{ov}}, t_{\text{grp}}) \cap A_g \neq \varnothing \}.$$

That is, any group with at least one discovery, is considered to be a selected group. Ideally, for any choice $(t_{\text{ov}}, t_{\text{grp}})$, we would like to be able to measure the overall false discovery proportion (FDP) at these thresholds,

$$\text{FDP}_{\text{ov}} = \text{FDP}_{\text{ov}}(t_{\text{ov}}, t_{\text{grp}}) = \frac{|\mathcal{H}^0 \cap \widehat{S}(t_{\text{ov}}, t_{\text{grp}})|}{1 \vee \left|\widehat{S}(t_{\text{ov}}, t_{\text{grp}})\right|} , \tag{3}$$

and the group FDP,

$$\text{FDP}_{\text{grp}} = \text{FDP}_{\text{grp}}(t_{\text{ov}}, t_{\text{grp}}) = \frac{|\mathcal{H}^0_{\text{grp}} \cap \widehat{S}_{\text{grp}}(t_{\text{ov}}, t_{\text{grp}})|}{1 \vee \left|\widehat{S}_{\text{grp}}(t_{\text{ov}}, t_{\text{grp}})\right|} . \tag{4}$$

To estimate these quantities, we define estimated overall FDP as

$$\widehat{\text{FDP}}_{\text{ov}} = \widehat{\text{FDP}}_{\text{ov}}(t_{\text{ov}}, t_{\text{grp}}) = \frac{n \cdot t_{\text{ov}}}{1 \vee \left|\widehat{S}(t_{\text{ov}}, t_{\text{grp}})\right|} ,$$

and the estimated group FDP as

$$\widehat{\text{FDP}}_{\text{grp}} = \widehat{\text{FDP}}_{\text{grp}}(t_{\text{ov}}, t_{\text{grp}}) = \frac{G \cdot t_{\text{grp}}}{1 \vee \left|\widehat{S}_{\text{grp}}(t_{\text{ov}}, t_{\text{grp}})\right|} .$$

We use the "hats" in our estimated FDP notation, to remind the reader that these quantities are empirical; we can explicitly calculate them from the data $P$ since they do not depend on knowing the underlying true set of nulls $\mathcal{H}^0$.

To understand these definitions, note that if there are $|\mathcal{H}^0|$ many null p-values which are uniformly distributed, then we expect roughly $|\mathcal{H}^0| \cdot t_{\text{ov}} \leq n \cdot t_{\text{ov}}$ many of them to lie below the threshold $t_{\text{ov}}$, and similarly for the $|\mathcal{H}^0_{\text{grp}}| \leq G$ many null groups. Therefore the numerators in $\widehat{\text{FDP}}_{\text{ov}}(t_{\text{ov}}, t_{\text{grp}})$ and $\widehat{\text{FDP}}_{\text{grp}}(t_{\text{ov}}, t_{\text{grp}})$ give intuitive (over)estimates of the numerators in the true false discovery proportions $\text{FDP}_{\text{ov}}(t_{\text{ov}}, t_{\text{grp}})$ and $\text{FDP}_{\text{grp}}(t_{\text{ov}}, t_{\text{grp}})$, respectively. (This is the motivation underlying the Benjamini-Hochberg procedure, extended also to the group setting.)

For any target FDR bounds $(\alpha_{\text{ov}}, \alpha_{\text{grp}})$, define the set of admissible thresholds

$$\widehat{\mathcal{T}}(\alpha_{\text{ov}}, \alpha_{\text{grp}}) = \left\{ (t_{\text{ov}}, t_{\text{grp}}) \in [0,1] \times [0,1] \ : \ \widehat{\text{FDP}}_{\text{ov}} \leq \alpha_{\text{ov}} \text{ and } \widehat{\text{FDP}}_{\text{grp}} \leq \alpha_{\text{grp}} \right\} .$$

Our first result shows that the set $\widehat{\mathcal{T}}(\alpha_{\text{ov}}, \alpha_{\text{grp}}) \subseteq [0,1] \times [0,1]$ has a well-defined maximum.

**Theorem 1.** *Fix any* $\alpha_{\text{ov}}, \alpha_{\text{grp}} \in [0,1]$ *and any vector of p-values* $P \in [0,1]^n$*. Define*

$$\widehat{t}_{\text{ov}} = \max\left\{t_{\text{ov}} \in [0,1] \ : \ \exists t_{\text{grp}} \in [0,1] \text{ s.t. } (t_{\text{ov}}, t_{\text{grp}}) \in \widehat{\mathcal{T}}(\alpha_{\text{ov}}, \alpha_{\text{grp}})\right\}, \textit{ and}$$
$$\widehat{t}_{\text{grp}} = \max\left\{t_{\text{grp}} \in [0,1] \ : \ \exists t_{\text{ov}} \in [0,1] \text{ s.t. } (t_{\text{ov}}, t_{\text{grp}}) \in \widehat{\mathcal{T}}(\alpha_{\text{ov}}, \alpha_{\text{grp}})\right\}.$$

*Then* $(\widehat{t}_{\text{ov}}, \widehat{t}_{\text{grp}}) \in \widehat{\mathcal{T}}(\alpha_{\text{ov}}, \alpha_{\text{grp}})$.

Intuitively, this result implies that $\widehat{\mathcal{T}}(\alpha_{\text{ov}}, \alpha_{\text{grp}})$ is a region in $[0,1] \times [0,1]$ that has a maximum "corner": a point $(\widehat{t}_{\text{ov}}, \widehat{t}_{\text{grp}})$ such that $(t_{\text{ov}}, t_{\text{grp}}) \leq (\widehat{t}_{\text{ov}}, \widehat{t}_{\text{grp}})$ for all points $(t_{\text{ov}}, t_{\text{grp}}) \in \widehat{\mathcal{T}}(\alpha_{\text{ov}}, \alpha_{\text{grp}})$. We remark that $\widehat{t}_{\text{ov}}$ and $\widehat{t}_{\text{grp}}$ always take values in a discrete grid,

$$\widehat{t}_{\text{ov}} \in \left\{\alpha_{\text{ov}} \cdot \frac{k}{n} : k = 0, \dots, n\right\} \text{ and } \widehat{t}_{\text{grp}} \in \left\{\alpha_{\text{grp}} \cdot \frac{k}{G} : k = 0, \dots, G\right\}.$$

The construction given in the above theorem defines our procedure: the p-filter procedure, applied to the given p-values $P$ and given partition into groups, returns the set of rejections/discoveries given by $\widehat{S}(\widehat{t}_{\text{ov}}, \widehat{t}_{\text{grp}})$. Recall that the set of discoveries $\widehat{S}(\widehat{t}_{\text{ov}}, \widehat{t}_{\text{grp}})$ consists of all hypotheses whose individual p-value $P_i$ and group p-value $\text{Simes}(P_{A_{g(i)}})$ *both* lie below their respective adaptive thresholds; the name "p-filter" refers to this process, where the rejected p-values are those that pass through both an individual-level filter and a group-level filter.

With our method now defined, we turn to a theorem on FDR control at both the individual and group level. First, we introduce the PRDS assumption, originally formulated by [3]:[1]

$$\begin{gathered}\text{For any nondecreasing set } D \subseteq [0,1]^n \text{ and any } i \in \mathcal{H}^0, \\ t \mapsto \mathbb{P}\left\{P \in D \mid P_i \leq t\right\} \text{ is a nondecreasing function over } t \in (0,1].\end{gathered} \tag{5}$$

We also assume that each true null p-value is uniformly distributed—in fact, our assumption is more flexible:

$$\text{For any } i \in \mathcal{H}^0, \mathbb{P}\left\{P_i \leq t\right\} \leq t \text{ for all } t \in [0,1]. \tag{6}$$

This assumption holds trivially if $P_i \sim \text{Uniform}[0,1]$, but also allows for a misspecified null distribution in some settings, or a discrete-valued p-value. We are now ready to state our result:

**Theorem 2.** *Let the p-values $P$ satisfy the assumptions* (5) *and* (6) *above, let* $(\widehat{t}_{\text{ov}}, \widehat{t}_{\text{grp}})$ *be defined as in Theorem 1, and let* $\widehat{S}(\widehat{t}_{\text{ov}}, \widehat{t}_{\text{grp}})$ *be the set of discoveries returned by the p-filter, as defined in* (2)*. Then the p-filter controls both overall and group FDR, i.e.*

$$\mathbb{E}\left[\text{FDP}_{\text{ov}}(\widehat{t}_{\text{ov}}, \widehat{t}_{\text{grp}})\right] \leq \alpha_{\text{ov}} \cdot \frac{\left|\mathcal{H}^0\right|}{n} \text{ and } \mathbb{E}\left[\text{FDP}_{\text{grp}}(\widehat{t}_{\text{ov}}, \widehat{t}_{\text{grp}})\right] \leq \alpha_{\text{grp}} \cdot \frac{\left|\mathcal{H}^0_{\text{grp}}\right|}{G}.$$

In fact, the setup described here is a special case of a multi-layer FDR framework that we describe below, where we seek to control FDR simultaneously across multiple partitions or partitions of the hypotheses. First, however, we describe an existing approach to the grouped FDR problem to compare it to our method for this setting.

[1] In [3]'s work, the assumption involves the probability $\mathbb{P}\left\{P \in D \mid P_i = t\right\}$, rather than conditioning on the event $P_i \leq t$ as in (5) which we prefer for later convenience; however, as discussed in their work, the two assumptions are equivalent. Also, recall that a set $D \in \mathbb{R}^n$ is called nondecreasing if $x \in D$ implies that $y \in D$ for any $y \geq x$ in the orthant ordering (i.e. $y \geq x$ if $y_i \geq x_i$ for all $i$).

## 3.2 Existing work: within-group FDR and group-level FDR

In recent work, [1] propose a related method for the multiple hypothesis testing problem with grouped structure. In their method, the first step is a screening step to select a set of groups of interest, $\widehat{S}_{\text{grp}}$; the mechanism for this screening step is determined by the user subject to some mild conditions. The second step is then to test the p-values within each selected group: for each $g \in \widehat{S}_{\text{grp}}$, run a selection procedure that controls the FDR at the level $\alpha_{\text{ov}} \cdot \frac{|\widehat{S}_{\text{grp}}|}{G}$. [10] develops this method further by examining a specific choice for the screening step:

1. First, apply the BH procedure with threshold $\alpha_{\text{grp}}$ to the Simes p-values of the $G$ groups,

$$\text{Simes}(P_{A_1}), \ldots, \text{Simes}(P_{A_G}),$$

   to select a set of groups $\widehat{S}_{\text{grp}}$. The group-level FDP is now given by $\frac{|\mathcal{H}^0_{\text{grp}} \cap \widehat{S}_{\text{grp}}|}{1 \vee |\widehat{S}_{\text{grp}}|}$.

2. Next, for each selected group $g \in \widehat{S}_{\text{grp}}$, run the BH procedure with threshold $\alpha_{\text{ov}} \cdot \frac{|\widehat{S}_{\text{grp}}|}{G}$ on the p-values within the group, $P_{A_g}$. Let $\widehat{S}_g$ be the selected set within group $g$. The FDP within group $g$ is now given by $\frac{|\mathcal{H}^0 \cap \widehat{S}_g|}{1 \vee |\widehat{S}_g|}$.

[10] show that the first step ensures that the group-level FDR is controlled at level $\alpha_{\text{grp}}$,

$$\mathbb{E}\left[\frac{|\mathcal{H}^0_{\text{grp}} \cap \widehat{S}_{\text{grp}}|}{1 \vee |\widehat{S}_{\text{grp}}|}\right] \leq \alpha_{\text{grp}}.$$

(This follows from the properties of Simes' test and BH procedure.) Furthermore, [1]'s results guarantee that the resulting average FDP across all selected groups is controlled as

$$\mathbb{E}\left[\frac{\sum_{g \in \widehat{S}_{\text{grp}}} (\text{FDP in group } g)}{1 \vee |\widehat{S}_{\text{grp}}|}\right] = \mathbb{E}\left[\frac{\sum_{g \in \widehat{S}_{\text{grp}}} \frac{|\mathcal{H}^0 \cap \widehat{S}_g|}{1 \vee |\widehat{S}_g|}}{1 \vee |\widehat{S}_{\text{grp}}|}\right] \leq \alpha_{\text{ov}}, \tag{7}$$

under the assumption that p-values in one group are independent of the other groups (with positive dependence allowed within each group).

Our p-filter method clearly has much in common with this procedure, but the two offer different types of guarantees. The p-filter does not offer control of the averaged within-group FDR; our guarantee is different, giving overall FDR control across all hypotheses selected. Depending on the setting, one or the other measure of false discovery control may be more desirable. We also note that the p-filter extends to a more general setting, discussed next, and is unique in allowing us to move to multiple partitions which are not necessarily arranged hierarchically, and allows dependence among p-values across groups.

# 4 Multilayer FDR control

We now turn to the more general problem of *multi-layer* FDR control, where we seek to control the false discovery rate across a range of arbitrary partitions of the hypotheses.

Suppose that we are given $n$ p-values, $P_1, \ldots, P_n \in [0,1]$, with an unknown set of nulls $\mathcal{H}^0 \subseteq [n]$. Furthermore, suppose we have $M$ partitions ("layers") of interest, with the $m$th partition having $G_m$ groups:

$$A_1^m, \ldots, A_{G_m}^m \subseteq [n]$$

for $m = 1, \ldots, M$. To return to the example mentioned in Section 1, in a fMRI study with $V$ voxels and $S$ timepoints, we might consider three layers:

- Layer $m = 1$ considers every voxel and timepoint separately ($V \cdot S$ groups);
- Layer $m = 2$ considers each voxel across all timepoints ($V$ groups);
- Layer $m = 3$ considers each timepoints across all voxels within each of $R$ regions of interest (ROIs) ($S \cdot R$ groups).

Define the null set for the $m$th partition as

$$\mathcal{H}_m^0 = \left\{ g \in [G_m] : A_g^m \subseteq \mathcal{H}^0 \right\} ,$$

and given a set $\widehat{S} \subseteq [n]$ of rejections, we define the $m$th rejection set as

$$\widehat{S}_m = \left\{ g \in [G_m] : \widehat{S} \cap A_g^m \neq \varnothing \right\} .$$

In our running fMRI example, for instance, $\mathcal{H}_2^0$ is the set of voxels $v$ such that $(v, s)$ is a null across *all* timepoints $s$, while $\widehat{S}_2$ is the set of voxels $v$ for which $(v, s)$ is a discovery for *any* timepoint $s$. Given a selected set $\widehat{S}$, define the FDP for the $m$th partition as

$$\text{FDP}_m(\widehat{S}) = \frac{\left| \widehat{S}_m \cap \mathcal{H}_m^0 \right|}{1 \vee |\widehat{S}_m|} .$$

Now we describe the p-filter procedure for this more general setting. Consider any thresholds $(t_1, \ldots, t_M) \in [0,1]^M$. We let

$$\begin{aligned} \widehat{S}(t_1, \ldots, t_M) &= \cap_{m=1}^M \left( \cup_{g=1,\ldots,G_m : \text{Simes}(P_{A_g^m}) \le t_m} A_g^m \right) \\ &= \left\{ i : \text{ for all } m, \ \text{Simes}(P_{A^m_{g(m,i)}}) \le t_m \right\} , \end{aligned} \tag{8}$$

where $g(m, i)$ indexes the group that $P_i$ belongs to in the $m$th partition. That is, for each partition $m$ we take the union of all groups whose Simes p-value is $\le t_m$; by taking the intersection across all layers, we see that a p-value $P_i$ is selected if, at every layer $m$, its group $A^m_{g(m,i)}$ passes this test. Correspondingly, we have

$$\widehat{S}_m(t_1, \ldots, t_M) = \left\{ g \in [G_m] : \widehat{S}(t_1, \ldots, t_M) \cap A_g^m \neq \varnothing \right\} . \tag{9}$$

We then let

$$\text{FDP}_m(t_1, \ldots, t_M) = \frac{\left| \widehat{S}_m(t_1, \ldots, t_M) \cap \mathcal{H}_m^0 \right|}{1 \vee |\widehat{S}_m(t_1, \ldots, t_M)|} ,$$

and define estimated FDP as

$$\widehat{\text{FDP}}_m(t_1,\dots,t_M) = \frac{G_m \cdot t_m}{1 \vee |\widehat{S}_m(t_1,\dots,t_M)|} \ .$$

Now define

$$\widehat{\mathcal{T}}(\alpha_1,\dots,\alpha_M) = \left\{(t_1,\dots,t_M) \in [0,1]^M : \widehat{\text{FDP}}_m(t_1,\dots,t_M) \le \alpha_m \text{ for all } m\right\} \ .$$

The next result proves that Theorem 1 extends to this more general setting, meaning that the set $\widehat{\mathcal{T}}(\alpha_1,\dots,\alpha_m)$ does indeed have a well-defined maximum point, thus defining our method.

**Theorem 3.** *Fix any $\alpha_1,\dots,\alpha_M \in [0,1]$ and any vector of p-values $P \in [0,1]^n$. Define*

$$\widehat{t}_m = \max\left\{t_m : \exists t_1,\dots,t_{m-1},t_{m+1},\dots,t_M \text{ s.t. } (t_1,\dots,t_M) \in \widehat{\mathcal{T}}(\alpha_1,\dots,\alpha_M)\right\}$$

*for each $m = 1,\dots,M$. Then*

$$(\widehat{t}_1,\dots,\widehat{t}_M) \in \widehat{\mathcal{T}}(\alpha_1,\dots,\alpha_M) \ .$$

The p-filter then selects the set

$$\widehat{S}(\widehat{t}_1,\dots,\widehat{t}_M).$$

As before, we remark that these adaptive thresholds take values on a discrete grid, with

$$\widehat{t}_m \in \left\{\alpha_m \cdot \frac{k}{G_m} : k = 0,\dots,G_m\right\} \tag{10}$$

for each $m$, but it is possible to find $(\widehat{t}_1,\dots,\widehat{t}_M)$ efficiently and without exhaustive search over this grid; see our algorithm given in Section 5.

Next, our main theorem shows that FDR is controlled simultaneously for each partition, by our p-filter:

**Theorem 4.** *Let the p-values $P \in [0,1]^n$ satisfy assumptions* (5) *and* (6) *above, and let $(\widehat{t}_1,\dots,\widehat{t}_M)$ be defined as in Theorem 3. Then for each $m = 1,\dots,M$, the method controls FDR for the $m$th partition,*

$$\mathbb{E}\left[\text{FDP}_m(\widehat{t}_1,\dots,\widehat{t}_M)\right] \le \alpha_m \cdot \frac{\left|\mathcal{H}^0_m\right|}{G_m} \ .$$

Clearly, this is a generalization of the setting considered previously, where the overall FDR and the group FDR can be controlled by defining two partitions, one that splits $[n]$ into $n$ many singleton sets, and one that is defined by the group structure. Note that the theoretical results for the initial setting, Theorems 1 and 2, are simply special cases of the more general results, Theorems 3 and 4, respectively. Unlike the overall FDR/group FDR setting, however, in general the $M$ partitions do not need to be nested; they are not constrained to form a hierarchy of partitions.

The proof of the above theorems are deferred to Appendix A, but one of the main ingredients is a technical lemma that could be of broader interest, and hence we state it below. First, for convenience, we define some notation: since the ratio "$\frac{0}{0}$" often arises in FDR control results, we let

$$\frac{a}{b} \overset{\cdots}{=} \begin{cases} \frac{a}{b}, & \text{if } b \neq 0, \\ 0, & \text{if } a = b = 0, \\ \text{undefined}, & \text{otherwise.} \end{cases} \tag{11}$$

We will use this to define conditional probability when the conditioned event has probability zero, that is, for two events $A, B$,

$$\mathbb{P}\{A \mid B\} = \frac{\mathbb{P}\{A \cap B\}}{\mathbb{P}\{B\}} = \begin{cases} \text{(the usual definition)}, & \mathbb{P}\{B\} > 0, \\ 0, & \mathbb{P}\{B\} = 0. \end{cases}$$

Let $X$ be an arbitrary random variable and write $F(y) = \mathbb{P}\{X \leq y\}$ for the cumulative density function of $X$. Hence, it trivially follows that

$$\mathbb{E}\left[\frac{\mathbf{1}\{X \leq y\}}{F(y)}\right] \leq 1 \text{ for any fixed constant } y.$$

Our main lemma below states that the above also holds for certain random $Y$.

**Lemma 1.** *Let $X, Y \in \mathbb{R}$ be random variables satisfying the assumption that*

$$\textit{For any } y\textit{, the function } x \mapsto \mathbb{P}\{Y < y \mid X < x\} \textit{ is nondecreasing in } x. \tag{12}$$

*Then, we have*

$$\mathbb{E}\left[\frac{\mathbf{1}\{X \leq Y\}}{F(Y)}\right] \leq 1.$$

The proof of Lemma 1 is given in Appendix A. This lemma gives an immediate corollary allowing us to understand the interaction between a null p-value $P_i$ and any function of the vector of p-values. First notice that for any null p-value $P_i$, our super-uniformity assumption (6) can be restated as

$$\mathbb{E}\left[\frac{\mathbf{1}\{P_i \leq t\}}{t}\right] \leq 1 \text{ for any fixed threshold } t.$$

The following corollary states that the above continues to remain true for certain random thresholds.

**Corollary 1.** *Let $P_i$ be null, satisfying super-uniformity assumption* (6)*, and assume that $P \in [0,1]^n$ is PRDS with respect to $P_i$. Then, for any function $f : [0,1]^n \to [0,\infty)$ that is nonincreasing (with respect to the orthant ordering), we have*

$$\mathbb{E}\left[\frac{\mathbf{1}\{P_i \leq f(P)\}}{f(P)}\right] \leq 1.$$

*Proof of Corollary 1.* We apply Lemma 1 by setting $X = P_i$ and $Y = f(P)$. Fix any $y \in \mathbb{R}$, and define $D = \{p \in \mathbb{R}^n : f(p) < y\}$. Since $f$ is a nonincreasing function, this means that $D$ is a nondecreasing set. Therefore,

$$\mathbb{P}\{Y < y \mid X < x\} = \mathbb{P}\{P \in D \mid P_i < x\}$$

is a nondecreasing function of $x$, by the PRDS assumption (5).[2] Writing $F$ to be the cumulative distribution function of $X = P_i$ and applying Lemma 1,

$$1 \geq \mathbb{E}\left[\frac{\mathbf{1}\{X \leq Y\}}{F(Y)}\right] = \mathbb{E}\left[\frac{\mathbf{1}\{P_i \leq f(P)\}}{F(f(P))}\right] \geq \mathbb{E}\left[\frac{\mathbf{1}\{P_i \leq f(P)\}}{f(P)}\right],$$

where the last step holds because $F(f(P)) \leq f(P)$ always by assumption (6). □

We note that this result is an extension of the work of [3] for analyzing FDR control of the BH procedure under the PRDS assumption; as part of their work, they prove an analogous result for the specific function $f(P) = \alpha \cdot \frac{\widehat{k}_\alpha(P)}{n}$ under the assumption $P_i \sim \text{Uniform}[0,1]$.

[2]While the PRDS assumption is stated using $\mathbb{P}\{P \in D \mid P_i \leq x\}$, we can replace this with conditioning on *strict* inequality by taking limits.

**Algorithm 1** The p-filter for multi-layer FDR control

**Input:** A vector of p-values $P \in [0,1]^n$; target FDR levels $\alpha_1, \ldots, \alpha_M$;
partition $m$ given by $A_1^m, \ldots, A_{G_m}^m \subseteq [n]$ for $m = 1, \ldots, M$.
**Initialize:** Thresholds $t_1 = \alpha_1, \ldots, t_M = \alpha_M$.
**repeat**
**for** $m = 1, \ldots, M$ **do**
Update the $m$th threshold: defining $\widehat{S}_m(\cdot)$ as in (9), let

$$t_m \leftarrow \max\left\{T \in [0, t_m] : \frac{G_m \cdot T}{1 \vee \left|\widehat{S}_m(t_1, \ldots, t_{m-1}, T, t_{m+1}, \ldots, t_M)\right|} \leq \alpha_m\right\} \tag{13}$$

**end for**
**until** the thresholds $t_1, \ldots, t_M$ are all unchanged in the last round.
**Output:** Adaptive thresholds $\widehat{t}_1 = t_1, \ldots, \widehat{t}_M = t_M$.

### 4.1 Comments on power and precision of the p-filter

The following points are worthy of note. As mentioned earlier, running the p-filter with one partition, which is the trivial finest partition, is exactly equivalent to running the classical BH procedure. Similarly, running the p-filter with two partitions, the finest one with threshold $\alpha_1$, and any other partition with threshold $\alpha_2$, is exactly equivalent to running the BH procedure if we set $\alpha_2 = \infty$. This observation can be further generalized to the case of $M$ partitions: running the p-filter with the first partition being the trivial one and with $\alpha_2 = \alpha_3 = ... = \alpha_M = \infty$ is exactly equivalent to running the BH procedure with $\alpha = \alpha_1$.

Since the set of discoveries is nondecreasing as a function of the thresholds $\alpha_1, \ldots, \alpha_M$, running the p-filter with nontrivial (i.e. finite) $\alpha_1, \ldots, \alpha_M$ leads to a set of discoveries that is no larger than the set produced by the BH procedure with threshold $\alpha = \alpha_1$; often the set is strictly smaller, and so p-filter's power is strictly lower. At the same time, we may often have lower *achieved* FDR as well, even at the individual level (the overall FDR), since the added layers of the p-filter can increase the precision of our discoveries.

As a simple example, consider a two layer partition with $n$ groups of size 1 at level $\alpha_1$, and with one group of size $n$ at level $\alpha_2$. We compare to BH with $\alpha = \alpha_1$ (equivalent to setting $\alpha_2 = \infty$). Then under the global null, if all p-values are independent and uniform, the probability of at least one rejection is equal to $\alpha_1$ for BH, and is equal to $\min\{\alpha_1, \alpha_2\}$ for the p-filter; under the global null this probability is equal to the FDR, so we see a lower *achieved* FDR for the p-filter.

## 5 Algorithm

Here, we present an efficient algorithm for implementing our method, given in Algorithm 1, which yields the correct solution according to the following result:

**Theorem 5.** *The output of Algorithm 1 is the vector of thresholds $(\widehat{t}_1, \ldots, \widehat{t}_m)$ defined in Theorem 3.*

Next we assess the run time of this algorithm. First, by definition of the algorithm, the $t_m$'s cannot increase; therefore the sets $\widehat{S}_m(t_1, \ldots, t_M)$ cannot increase over the iterations of the algorithm, and so the denominator in step (13) is nonincreasing. Therefore, for each run of the outer loop (the "repeat...until" loop), either $t_m$ decreases strictly for some $m$, or all $t_m$'s stay the same and so the algorithm terminates. Furthermore, observe that the maximizer $t_m$ to the update step (13) must lie in the set $\left\{\frac{\alpha_m k}{G_m} : k = 0, \dots, G_m\right\}$. This means that there can be at most $G_1 + \cdots + G_M$ distinct instances where one of the $t_m$'s decreases, and so the algorithm terminates after at most $G_1 + \cdots + G_M + 1$ passes through the outer loop.

# 6 Experiments with simulated data

In this section we examine two designs: one setting where we seek to control individual-level and group-level FDR control as discussed in Section 3, and a second more complex setting where we consider three different partitions of the hypotheses simultaneously. We compare the p-filter with the BH method and with [1] method (denoted as "BB" throughout this section). For both experiments, all p-values in the simulations are independent and are generated as follows:

$$X \sim \mu + \mathcal{N}(0,1); \text{ p-value} = 1 - \Phi(X) \tag{14}$$

where $\Phi$ is the standard Gaussian CDF, with $\mu = 0$ for nulls and $\mu > 0$ for true signals. Larger values of $\mu$ correspond to stronger true signals that are easier to detect. All simulations were run in R [11]. [3]

## 6.1 Grouped setting

In our first simulation, we consider a simple grouped scenario: we have $n = 1,000$ hypotheses, partitioned into 100 groups of size 10. There are 55 true signals: one in group 1, two in group 2, ..., and ten in group 10.

Figure 2 shows the outcome of one trial run of the simulation (with $\mu = 3$); for convenience, we display our tests in a $10 \times 100$ array where each column corresponds to a group and the first 10 columns contain all the true signals. We see that the p-filter and the BB method both select very few null columns (groups), which is desirable; in fact, the results from these two methods are nearly identical. BH, which does not use the partition of the hypotheses, selects many null groups (columns), but is also slightly better able to find the true signals.

Results across a range of $\mu$ values are shown in Figure 3, plotting FDR and power for this array of hypotheses at the individual (entry-wise) and group (column-wise) levels. We see that the three methods have very similar power (with slightly higher power for BH), but different FDR control properties: while all three methods control entry-wise FDR, as expected we see that BH does not control column-wise FDR. The p-filter and BB methods show nearly identical results, with very slightly higher entry-wise power for BB.

## 6.2 Multilayer setting

We now consider a setting where the structure of the true signals is best captured using multiple partitions of the data. In this setting, the $n = 10,000$ hypotheses are arranged into a $100 \times 100$ grid.

[3] R code implementing our method through Algorithm 1, along with scripts reproducing all simulated experiments presented in this paper, can be found at `http://www.stat.uchicago.edu/~rina/pfilter.html`.

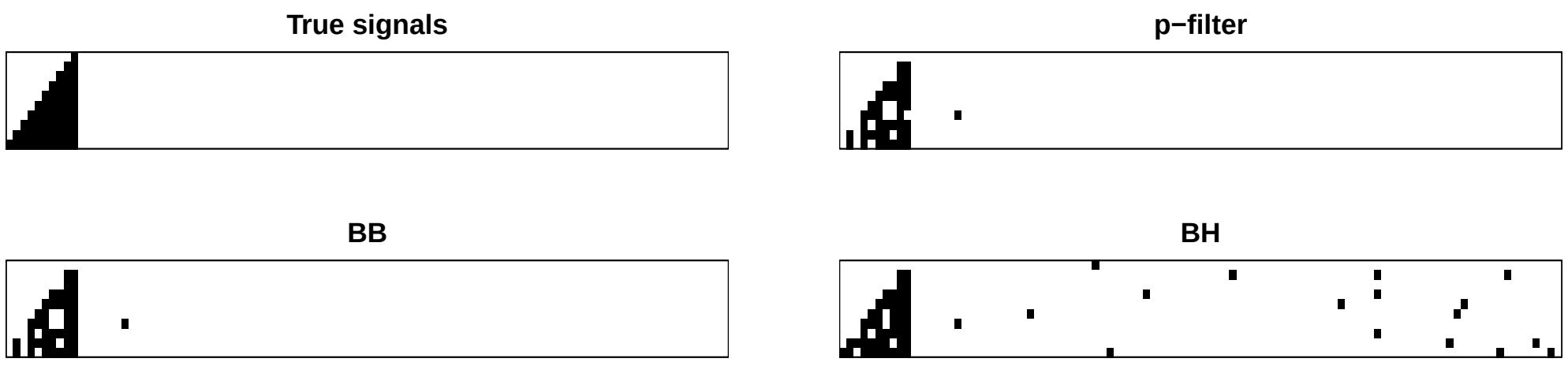

Figure 2: A demonstration of one trial run of the group-wise sparsity simulation (Section 6.1).

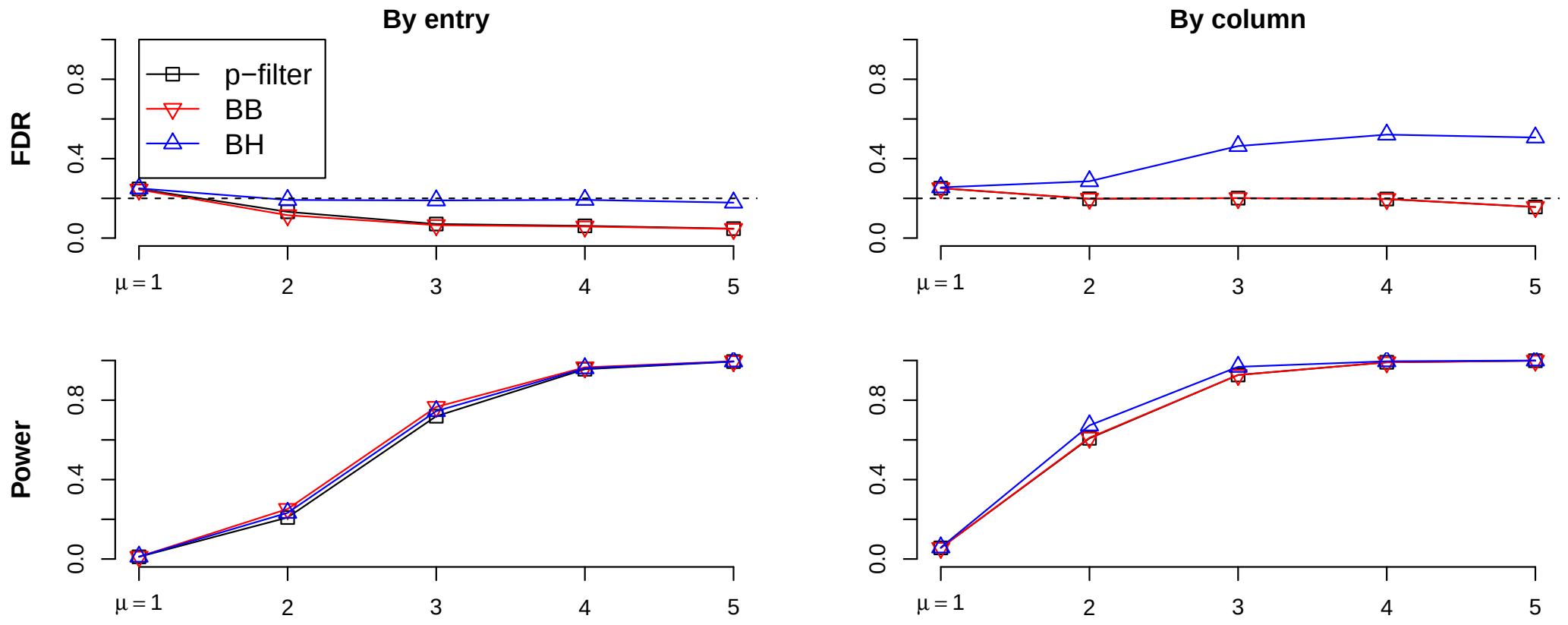

Figure 3: Results for the group-wise sparsity simulation (Section 6.1), averaged over 100 trials. The dotted lines show the target FDR level for each of the partitions.

The true signals lie in two $15 \times 15$ blocks, plus 15 additional signals that lie along a diagonal, and are therefore alone in their respective rows and columns (see the top-left block of Figure 4). Therefore, they are sparse at the individual (entry-wise) level, but also are row-wise sparse and column-wise sparse. The 15 signals along the diagonal make this simulation more challenging for the p-filter and BB methods, which are best able to find signals that are grouped together. We again compare three methods: the p-filter (with three layers: entries, rows, and columns); the BB procedure (where the groups are defined by the rows); and the BH procedure.

Figure 4 shows the outcome of one trial run of the simulation (with $\mu = 3$). We see that the p-filter selects few null rows and columns, which is desirable. BB, with the groups defined as rows, selects few null rows but many null columns. BH, which does not use row/column information, selects many null rows and columns. On the other hand, BH is much better able to find the sparse signals along the diagonal, as expected.

Results across a range of $\mu$ values are shown in Figure 5, plotting FDR and power at the entry-wise, row-wise, and column-wise levels for this two-dimensional array of hypotheses. At the entry-wise level, the three methods have similar power and all control FDR. For rows, BH does not control row-wise FDR as expected, but is able to achieve higher power (due to the sparse signals along the diagonal). For columns, BH and BB both lose FDR control as expected, with a corresponding slight increase in power. The p-filter controls all three forms of FDR, as guaranteed by our theoretical results, and achieves good power across the three layers.

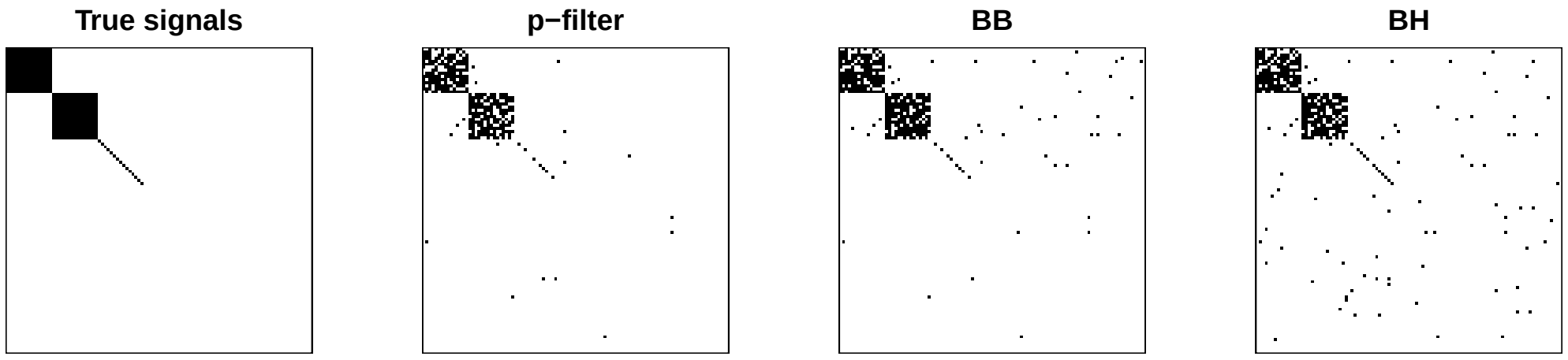

Figure 4: A demonstration of one trial of the row- and column-wise sparsity simulation (Section 6.2).

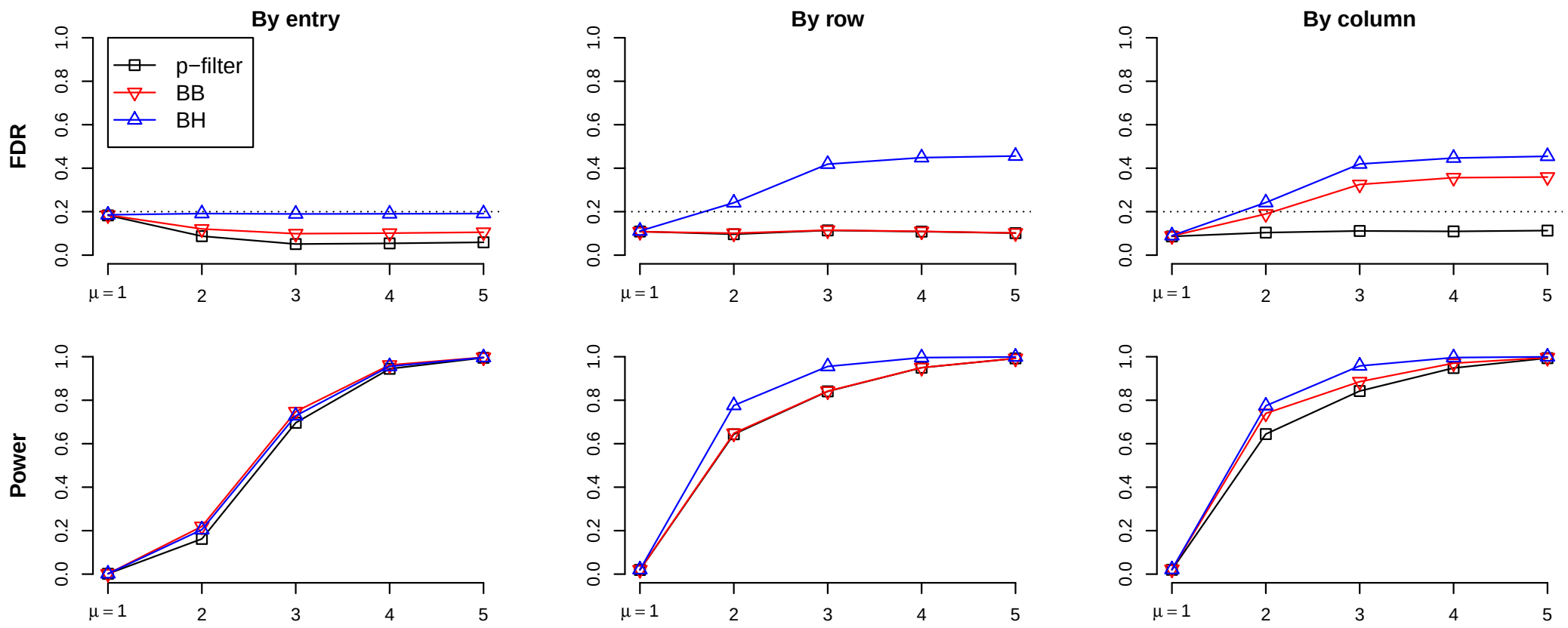

Figure 5: Results for the row- and column-wise sparsity simulation (Section 6.2), averaged over 100 trials. The dotted lines show the target FDR level for each of the partitions.

# 7 Experiment with fMRI data

We now demonstrate one way to use spatial and temporal prior information to aid inference in neuroscientific applications. We use freely available fMRI data from [14]. 8 subjects read a chapter of Harry Potter and the SorcererâĂŹs Stone while words were presented one at a time. The total presentation time is 2710 seconds and the available data consists of 1355 volumes of fMRI activity (one scan every 2 seconds) for each of the 8 subjects, each scanned with the same timeline of stimulus presentation. Each subject's brain is represented in (3 mm)×(3 mm)×(3 mm) voxels, which are all normalized to the same coordinate space, with 41,073 voxels common to all 8 subjects. The text was annotated with multiple types of intermediate features: in the analysis that follows, we use the semantic annotations available on the paper's accompanying website.

**Temporal Prior Information.** The fMRI machine measures the hemodynamic response, a delayed response that is the neural correlate of brain activity corresponding to changes in the magnetic field due to blood flowing into the brain as a result of brain activity. Every fMRI sample could therefore be approximated as the superposition of events happening in the 8 to 10 preceding seconds. It is hence appropriate to ask whether the features presented at time $t$ are able to predict the brain activity at time $t + s$, for $s = 4, 6, 8$ seconds, which based on prior knowledge corresponds to the peak of the hemodynamic response.

**Computing p-values for feature-activity correlations.** For each delay $s$, we use the same predictive encoding model as proposed in [14], where one fits a linear regression model from the text's semantic features to each voxel's recorded fMRI brain activity delayed by $s$ seconds. Data from all 8 subjects is used for this to boost the signal to noise ratio. It was determined in [15] that the obtained results and conclusions on this dataset are quite stable to various modeling and algorithmic choices like regularization and smoothing. Hence the exact methods used are not very relevant for our present purposes and the reader is directed to [14, 15] for more details. This finally yields a p-value $P_{v,s}$ for each voxel $v$ and delay $s$. Each p-value $P_{v,s}$ represents the question, "Is voxel $v$ correlated with the semantic features of the text presented $s$ seconds earlier?" Figure 6 displays these p-values on a brain (for $s = 6$ seconds, in negative logarithm scale), using the Pycortex software by [5]. The natural spatio-temporal correlation in the brain data, along with the "searchlight" procedure used in [14], results in a slightly smoothed set of positively correlated p-values, which we assume satisfy PRDS.

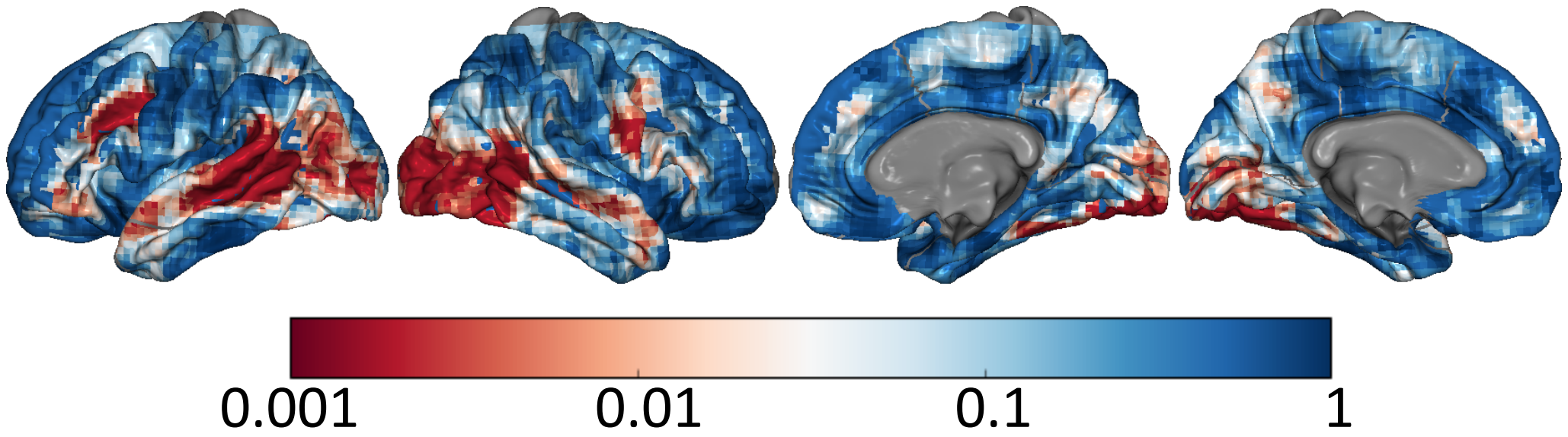

Figure 6: For time delay $s = 6$, original p-values (between 1 and $10^{-3}$) are plotted in negative log-scale, one for each voxel in the brain, are plotted on the outside (lateral, left) and inside (medial, right) of the brain. Red regions correspond to a high correlation between semantic features and brain activity $s = 6$ seconds after stimulus presentation, while blue regions correspond to very low correlation. (dark grey = no readings)

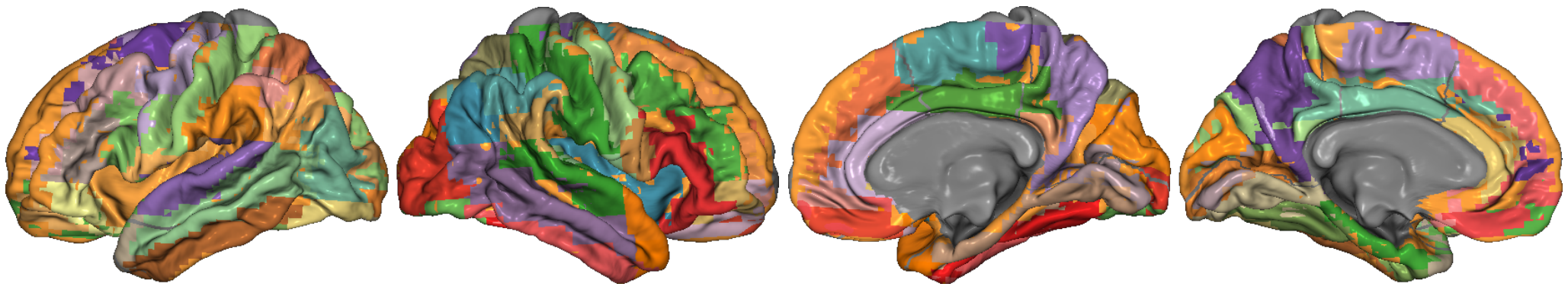

Figure 7: The 90 regions of interest of the brain as used by our experiments, each in a different color (the colors have no meaning, and are purely for easy visualization).

**Spatial prior information** Neuroscientists often divide the voxels into *regions of interest* (ROIs), which are intended to be functionally distinct areas of the brain, like the visual cortex, the hippocampus, the auditory cortex, etc. Figure 7 shows the 90 ROIs that we use in this paper, marked in different colors for easy visualization. While the exact number of ROIs and their precise bound-aries is still debated, these still provide reasonable prior information for contiguous regions of space where the activity may be correlated with the input stimulus.

**Applying the p-filter** We provide 3 different non-hierarchically arranged partitions. The first is the trivial finest partition with $41073 \times 3$ individual p-values, denoted $P_{v,s}$ as before, for $v = 1, ..., 41073$ and $s = 4, 6, 8$. The second partition uses temporal information to group $P_{v,4}, P_{v,6}$ and $P_{v,8}$ together for each $v$ (41073 many groups). The third partition uses spatial information to group together $P_{v,s}$ for all voxels $v$ in the same ROI, for each $s$ (90×3 many groups). We set $\alpha_1 = 0.05, \alpha_2 = 0.05, \alpha_3 = 0.1$. For $s = 6$, Figure 8 displays the rejected p-values in red, and the non-rejected p-values in grey.

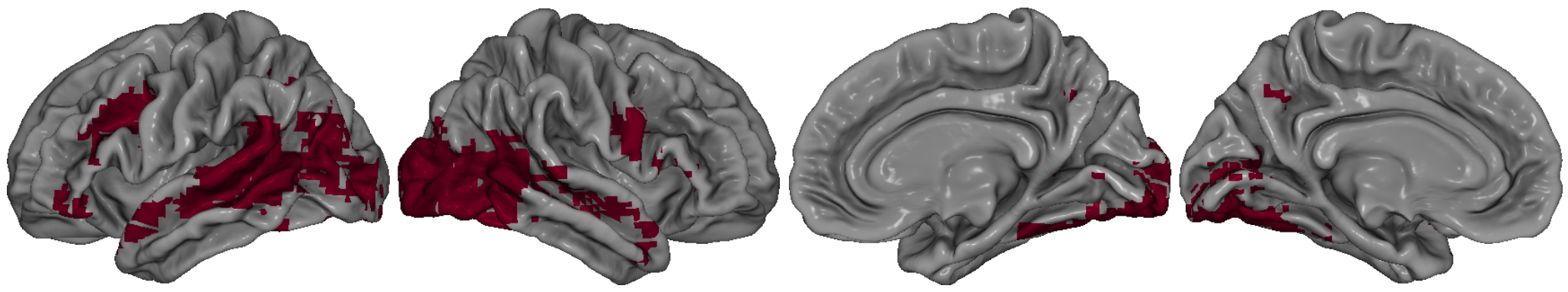

Figure 8: For time delay $s = 6$, we display the final results obtained by the p-filter method, with discoveries marked in dark red and non-discoveries in light grey.

The ground truth is, of course, unknown, and this example serves as one possible way to construct layers and analyze the given brain data. It is now a fairly standard procedure in neuroscience to use BH (in this case, directly on the input $41073 \times 3$ p-values) — recall that this just corresponds to a special case of our p-filter procedure, one that does not explicitly take temporal or spatial structure into account. As mentioned earlier, when used to control FDR at both the individual voxel and group levels, our procedure may have lower power than the usual BH procedure, since p-values must pass individual *and* group-level constraints; however, as demonstrated in the earlier simulations, these constraints often help achieve *nearly* the same power but with a sizable reduction in the achieved number of false discoveries, resulting in an improved *precision*. This may allow the scientist to possibly employ higher FDR thresholds, as has been recognized as important for fMRI data by [8].

# 8 Conclusion

We introduced an extremely flexible method, the p-filter, that simultaneously controls the false discovery rate (FDR) across multiple layers (partitions of p-values), a guarantee that is significantly more general than existing work. We gave an efficient algorithm for computing the set of discoveries (i.e., rejected p-values), given all the p-values, their various partitions, and a target FDR for each partition. We demonstrated its usefulness in simulations—when the pattern of true signals was naturally grouped across rows and columns, we applied the p-filter for entry-, row-, and column-wise FDR control, and achieved higher precision, i.e., nearly the same power at lower FDR. We conjecture that this approach may find widespread usage in spatio-temporal or other multimodal applications where p-values can naturally be grouped in many ways across modalities.

**Acknowledgments** The authors would like to thank the American Institute of Mathematics (AIM)'s Workshop on Inference in High-Dimensional Regression, where this collaboration started. The authors are also very grateful to Leila Wehbe, who generously shared her time, plotting tools and fMRI data.

# A Proofs

## A.1 Proof of Theorem 3

For each $m$, by definition of $\widehat{t}_m$, there is some $t^m_1, \dots, t^m_{m-1}, t^m_{m+1}, \dots, t^m_M$ such that

$$(t^m_1, \dots, t^m_{m-1}, \widehat{t}_m, t^m_{m+1}, \dots, t^m_M) \in \widehat{\mathcal{T}}(\alpha_1, \dots, \alpha_M) \ . \tag{15}$$

Thus, for each $m' \neq m$, $\widehat{t}_{m'} \geq t^m_{m'}$ by definition of $\widehat{t}_{m'}$. Then

$$\widehat{S}(t^m_1, \dots, t^m_{m-1}, \widehat{t}_m, t^m_{m+1}, \dots, t^m_M) \subseteq \widehat{S}(\widehat{t}_1, \dots, \widehat{t}_{m-1}, \widehat{t}_m, \widehat{t}_{m+1}, \dots, \widehat{t}_M) \ ,$$

because $\widehat{S}(t_1,\dots,t_M)$ is a nondecreasing function of $(t_1,\dots,t_M)$. Therefore,

$$\widehat{\text{FDP}}_m(\widehat{t}_1,\dots,\widehat{t}_{m-1},\widehat{t}_m,\widehat{t}_{m+1},\dots,\widehat{t}_M) = \frac{G_m\cdot\widehat{t}_m}{1\vee\left|\widehat{S}_m(\widehat{t}_1,\dots,\widehat{t}_{m-1},\widehat{t}_m,\widehat{t}_{m+1},\dots,\widehat{t}_M)\right|}$$
$$\le \frac{G_m\cdot\widehat{t}_m}{1\vee\left|\widehat{S}_m(t^m_1,\dots,t^m_{m-1},\widehat{t}_m,t^m_{m+1},\dots,t^m_M)\right|}\le\alpha_m\ ,$$

where the last step holds by definition of $\widehat{\mathcal{T}}(\alpha_1,\dots,\alpha_M)$ and uses Eq. (15). Since this holds for all $m$, this proves that $(\widehat{t}_1,\dots,\widehat{t}_M)\in\widehat{\mathcal{T}}(\alpha_1,\dots,\alpha_M)$ by definition of $\widehat{\mathcal{T}}(\alpha_1,\dots,\alpha_M)$.

## A.2 Proof of Theorem 4

Fix any partition $m$. Since $\mathbb{P}\{P_i=0\}=0$ for any $i\in\mathcal{H}^0$ by our assumption (6), we assume that $P_i\neq 0$ for any $i\in\mathcal{H}^0$ without further mention; this assumption then implies that if $g\in\widehat{S}_m(\widehat{t}_1,\dots,\widehat{t}_M)$ for some null group $g\in\mathcal{H}^0_m$, we must have $\widehat{t}_m>0$. We then calculate

$$\text{FDP}_m(\widehat{t}_1,\dots,\widehat{t}_M)=\frac{\left|\widehat{S}_m(\widehat{t}_1,\dots,\widehat{t}_M)\cap\mathcal{H}^0_m\right|}{1\vee\left|\widehat{S}_m(\widehat{t}_1,\dots,\widehat{t}_M)\right|}=\sum_{g\in\mathcal{H}^0_m}\frac{\mathbf{1}\left\{g\in\widehat{S}_m(\widehat{t}_1,\dots,\widehat{t}_M)\right\}}{1\vee\left|\widehat{S}_m(\widehat{t}_1,\dots,\widehat{t}_M)\right|}$$
$$\le\alpha_m\cdot\sum_{g\in\mathcal{H}^0_m}\frac{\mathbf{1}\left\{g\in\widehat{S}_m(\widehat{t}_1,\dots,\widehat{t}_M)\right\}}{\widehat{t}_mG_m},\tag{16}$$

since $\frac{\widehat{t}_mG_m}{1\vee|\widehat{S}_m(\widehat{t}_1,\dots,\widehat{t}_M)|}=\widehat{\text{FDP}}_m(\widehat{t}_1,\dots,\widehat{t}_M)\le\alpha_m$ by definition of the method. (The notation $\frac{a}{b}$ is defined in Eq. (11).)

Now fix any null group $g\in\mathcal{H}^0_m$. Define $\widehat{k}^m_g=\widehat{k}_{\widehat{t}_m}(P_{A^m_g})$, the number of rejections when group $A^m_g$ is tested with the BH procedure with threshold $\widehat{t}_m$. Then, by definition of $\widehat{S}$, if $A^m_g$ is rejected then we must have $\text{Simes}(P_{A^m_g})\le\widehat{t}_m$ and so, as argued in Eq. (1), $A^m_g$ passes the BH procedure at threshold $\widehat{t}_m$; that is,

$$g\in\widehat{S}_m(\widehat{t}_1,\dots,\widehat{t}_M)\ \Rightarrow\ \widehat{k}^m_g>0,$$

and this can only occur when $\widehat{t}_m>0$ since $P_i\neq0$ for all $i\in A^m_g\subseteq\mathcal{H}^0$. Furthermore,

$$\mathbf{1}\left\{\widehat{k}^m_g>0\right\}=\frac{\widehat{k}^m_g}{\widehat{k}^m_g}=\frac{\sum_{i\in A^m_g}\mathbf{1}\left\{P_i\le\frac{\widehat{t}_m\widehat{k}^m_g}{|A^m_g|}\right\}}{\widehat{k}^m_g}=\sum_{i\in A^m_g}\frac{\mathbf{1}\left\{P_i\le\frac{\widehat{t}_m\widehat{k}^m_g}{|A^m_g|}\right\}}{\widehat{k}^m_g}.$$

Therefore, for each $g\in\mathcal{H}^0_m$, we can write

$$\frac{\mathbf{1}\left\{g\in\widehat{S}_m(\widehat{t}_1,\dots,\widehat{t}_M)\right\}}{\widehat{t}_mG_m}\le\frac{\mathbf{1}\left\{\widehat{k}^m_g>0\right\}}{\widehat{t}_mG_m}=\frac{1}{G_m|A^m_g|}\sum_{i\in A^m_g}\frac{\mathbf{1}\left\{P_i\le\frac{\widehat{t}_m\widehat{k}^m_g}{|A^m_g|}\right\}}{\frac{\widehat{t}_m\widehat{k}^m_g}{|A^m_g|}}.$$

So, returning to Eq. (16), we conclude

$$\text{FDP}_m(\widehat{t}_1,\dots,\widehat{t}_M)\le\sum_{g\in\mathcal{H}^0_m}\frac{\alpha_m}{G_m|A^m_g|}\sum_{i\in A^m_g}\frac{\mathbf{1}\left\{P_i\le\frac{\widehat{t}_m\widehat{k}^m_g}{|A^m_g|}\right\}}{\frac{\widehat{t}_m\widehat{k}^m_g}{|A^m_g|}}.$$

Next, let $f^m_g:[0,1]^n\to[0,1]$ be the function that maps $P$ to $\frac{\widehat{t}_m\widehat{k}^m_g}{|A^m_g|}$. We observe that

- $\widehat{t}_m$ is a nonincreasing function of $P$ by definition of our procedure; and
- $\widehat{k}^m_g$ is also nonincreasing in $P$: if $P$ is lower, then the threshold $\widehat{t}_m$ can only rise; lower p-values and a higher (less conservative) threshold can only increase the number of rejections.

Hence $f^m_g$ is a nonincreasing function of $P$. By Corollary 1, $\mathbb{E}\left[\frac{\mathbf{1}\left\{P_i\le f^m_g(P)\right\}}{f^m_g(P)}\right]\le1$, thus

$$\mathbb{E}\left[\text{FDP}_m(\widehat{t}_1,\dots,\widehat{t}_M)\right]\le\sum_{g\in\mathcal{H}^0_m}\frac{\alpha_m}{G_m|A^m_g|}\sum_{i\in A^m_g}(1)=\sum_{g\in\mathcal{H}^0_m}\frac{\alpha_m}{G_m}=\alpha_m\frac{|\mathcal{H}^0_m|}{G_m}.$$

## A.3 Proof of Theorem 5

First we introduce some notation: let $(t_1^{(k)},\dots,t_M^{(k)})$ be the thresholds after the $k$th pass through the algorithm. We prove that $t_m^{(k)} \geq \widehat{t}_m$ for all $m,k$, by induction. At initialization, $t_m^{(0)} = \alpha_m \geq \widehat{t}_m$ for all $m$. Now suppose that $t_m^{(k-1)} \geq \widehat{t}_m$ for all $m$; we now show that $t_m^{(k)} \geq \widehat{t}_m$ for all $m$.

To do this, consider the $m$th "layer" of the $k$th pass through the algorithm. Before this stage, we have thresholds $t_1^{(k)},\dots,t_{m-1}^{(k)},t_m^{(k-1)},t_{m+1}^{(k-1)},\dots,t_M^{(k-1)}$, and we now update $t_m^{(k)}$. Applying induction also to this inner loop, assume that $t_{m'}^{(k)} \geq \widehat{t}_{m'}$ for all $m'=1,\dots,m-1$. We now prove that $t_m^{(k)} \geq \widehat{t}_m$. By definition,

$$t_m^{(k)} = \max\left\{T : \frac{G_m \cdot T}{1 \vee \left|\widehat{S}_m(t_1^{(k)},\dots,t_{m-1}^{(k)},T,t_{m+1}^{(k-1)},\dots,t_M^{(k-1)})\right|} \leq \alpha_m\right\}. \tag{17}$$

Since $t_{m'}^{(k)} \geq \widehat{t}_{m'}$ for all $m'=1,\dots,m-1$, and $t_{m'}^{(k-1)} \geq \widehat{t}_{m'}$ for all $m'=m+1,\dots,M$,

$$\frac{G_m \cdot \widehat{t}_m}{1 \vee \left|\widehat{S}_m(t_1^{(k)},\dots,t_{m-1}^{(k)},\widehat{t}_m,t_{m+1}^{(k-1)},\dots,t_M^{(k-1)})\right|} \leq \frac{G_m \cdot \widehat{t}_m}{1 \vee \left|\widehat{S}_m(\widehat{t}_1,\dots,\widehat{t}_{m-1},\widehat{t}_m,\widehat{t}_{m+1},\dots,\widehat{t}_M)\right|}$$

which is $\leq \alpha_m$ by definition of $(\widehat{t}_1,\dots,\widehat{t}_M)$. Therefore, $\widehat{t}_m$ is in the feasible set for Eq. (17), and so we must have $t_m^{(k)} \geq \widehat{t}_m$. By induction this is then true for all $k,m$.

Now suppose that the algorithm stabilizes at thresholds $(t_1^{(k)},\dots,t_M^{(k)})$, after $k$ passes through the algorithm. After completing the $m$th layer of the last pass through the algorithm, we have thresholds $t_1^{(k)},\dots,t_m^{(k)},t_{m+1}^{(k-1)},\dots,t_M^{(k-1)}$; however, since the algorithm stops after the $k$th pass, this means that $t_{m'}^{(k-1)} = t_{m'}^{(k)}$ for all $m'$. By definition of $t_m^{(k)}$,

$$\frac{G_m \cdot t_m^{(k)}}{1 \vee \left|\widehat{S}_m(t_1^{(k)},\dots,t_{m-1}^{(k)},t_m^{(k)},t_{m+1}^{(k)},\dots,t_M^{(k)})\right|} \leq \alpha_m.$$

This means that $(t_1^{(k)},\dots,t_M^{(k)}) \in \widehat{\mathcal{T}}(\alpha_1,\dots,\alpha_M)$, and so $t_m^{(k)} \leq \widehat{t}_m$ by Theorem 3. But by the work above, we also know that $t_m^{(k)} \geq \widehat{t}_m$; this proves the theorem.

## A.4 Proof of Lemma 1

Fix any $\epsilon > 0$. Recalling that $F$ is the CDF of $X$, we define a sequence $+\infty = y_0 > y_1 > y_2 > \dots$ as follows: for each $i \geq 0$ define

$$y_{i+1} := \min\left\{y : F(y) \geq \frac{F_-(y_i)}{1+\epsilon}\right\},$$

where $F_-(y) := \sup\{F(y') : y' < y\} = \mathbb{P}\{X < y\}$. Trivially, $\lim_{i\to\infty} F(y_i) = 0$, so

$$\{y \in \mathbb{R} : F(y) > 0\} = \cup_{i\geq 0}[y_{i+1},y_i). \tag{18}$$

Therefore, it follows that

$$\begin{aligned}
\mathbb{E}\left[\frac{\mathbf{1}\{X \leq Y\}}{F(Y)}\right] &= \mathbb{E}\left[\frac{\mathbf{1}\{X \leq Y\}}{F(Y)} \cdot \sum_{i\geq 0}\mathbf{1}\{y_{i+1} \leq Y < y_i\}\right] \quad \text{by (18)}\\
&\leq \sum_{i\geq 0}\mathbb{E}\left[\frac{\mathbf{1}\{X < y_i\}}{F(y_{i+1})} \cdot \mathbf{1}\{y_{i+1} \leq Y < y_i\}\right]\\
&\leq (1+\epsilon)\cdot\sum_{i\geq 0}\mathbb{E}\left[\frac{\mathbf{1}\{X < y_i\}}{F_-(y_i)} \cdot \mathbf{1}\{y_{i+1} \leq Y < y_i\}\right] \quad \text{by definition of } y_{i+1}.
\end{aligned}$$

Now define the following partial sum for any $n \geq m \geq 0$ :

$$S_{m,n} = \sum_{i=m}^{n}\mathbb{E}\left[\frac{\mathbf{1}\{X < y_i\}}{F_-(y_i)} \cdot \mathbf{1}\{y_{i+1} \leq Y < y_i\}\right].$$

We claim that

$$S_{m,n} \leq \mathbb{P}\{Y < y_m \mid X < y_m\} \text{ for all } n \geq m \geq 0. \tag{19}$$

Assuming for the moment that the above claim is true, we have

$$\mathbb{E}\left[\frac{\mathbf{1}\{X < Y\}}{F(Y)}\right] \leq (1+\epsilon) \cdot \sum_{i \geq 0} \mathbb{E}\left[\frac{\mathbf{1}\{X < y_i\}}{F_-(y_i)} \cdot \mathbf{1}\{y_{i+1} \leq Y < y_i\}\right] = (1+\epsilon) \cdot \lim_{n \to \infty} S_{0,n},$$

where the limit holds since we have an infinite sum of nonnegative terms. Since $\epsilon > 0$ is arbitrarily small and Eq. (19) implies $S_{0,n} \leq 1$, this proves $\mathbb{E}\left[\frac{\mathbf{1}\{X<Y\}}{F(Y)}\right] \leq 1$ as desired.

It remains to be shown that Eq. (19) holds for all $n \geq m \geq 0$. We prove this for each fixed $n$ by induction over $m$. Starting with $m = n$, the bound is true trivially. Assuming it's true for some $m \geq 1$, we next prove it with $m - 1$ in place of $m$. We have

$$\begin{aligned}
S_{m-1,n} &= \mathbb{E}\left[\frac{\mathbf{1}\{X < y_{m-1}\}}{F_-(y_{m-1})} \cdot \mathbf{1}\{y_m \leq Y < y_{m-1}\}\right] + S_{m,n} \quad \text{by definition} \\
&\leq \mathbb{E}\left[\frac{\mathbf{1}\{X < y_{m-1}\}}{F_-(y_{m-1})} \cdot \mathbf{1}\{y_m \leq Y < y_{m-1}\}\right] + \mathbb{P}\{Y < y_m \mid X < y_m\} \quad \text{by (19)} \\
&\leq \mathbb{E}\left[\frac{\mathbf{1}\{X < y_{m-1}\}}{F_-(y_{m-1})} \cdot \mathbf{1}\{y_m \leq Y < y_{m-1}\}\right] + \mathbb{P}\{Y < y_m \mid X < y_{m-1}\} \quad \text{by (12)} \\
&= \mathbb{P}\{y_m \leq Y < y_{m-1} \mid X < y_{m-1}\} + \mathbb{P}\{Y < y_m \mid X < y_{m-1}\} \\
&= \mathbb{P}\{Y < y_{m-1} \mid X < y_{m-1}\},
\end{aligned}$$

proving that (19) holds with $m - 1$ in place of $m$. This concludes the proof.